# Multiple Symmetry-Protected Dirac Nodal Lines in A Quasi-One-Dimensional Semimetal

Zhanyang Hao[1,*], Weizhao Chen[1,*], Yuan Wang[1], Jiayu Li[1], Xiao-Ming Ma[1], Yu-Jie Hao[1], Ruie Lu[1], Zecheng Shen[1], Zhicheng Jiang[2,5], Wanling Liu[2], Qi Jiang[2], Xiao Lei[1,6], Le Wang[1], Ying Fu[1,7], Liang Zhou[1], Lianglong Huang[1], Zhengtai Liu[2], Mao Ye[2], Dawei Shen[2], Jiawei Mei[1], Hongtao He[1], Cai Liu[1], Ke Deng[1], Chang Liu[1], Qihang Liu[1,3,4,#], Chaoyu Chen[1,#]

[1] Shenzhen Institute for Quantum Science and Engineering (SIQSE) and Department of Physics, Southern University of Science and Technology (SUSTech), Shenzhen 518055, China.

[2] State Key Laboratory of Functional Materials for Informatics and Center for Excellence in Superconducting Electronics, Shanghai Institute of Microsystem and Information Technology, Chinese Academy of Sciences, Shanghai 200050, China.

[3] Guangdong Provincial Key Laboratory for Computational Science and Material Design, Southern University of Science and Technology, Shenzhen 518055, China.

[4] Shenzhen Key Laboratory of for Advanced Quantum Functional Materials and Devices, Southern University of Science and Technology, Shenzhen 518055, China.

[5] Center of Materials Science and Optoelectronics Engineering, University of Chinese Academy of Sciences, Beijing 100049, China.

[6] Department of Physics, The Hong Kong University of Science and Technology, Clear Water Bay, Hong Kong, China.

[7] Institute of Applied Physics and Materials Engineering, University of Macau, Avenida da Universidade Taipa, Macau 999078, P. R. China.

[*]These authors contributed equally to this work.

[#]Correspondence should be addressed to Q.L. (liuqh@sustech.edu.cn) and C.C. (chency@sustech.edu.cn)

**Abstract**

Nodal-line semimetals (NLSMs) contains Dirac/Weyl type band-crossing nodes extending into shapes of line, loop and chain in the reciprocal space, leading to novel band topology and transport responses. Robust NLSMs against spin-orbit coupling typically occur in three-dimensional materials with more symmetry operations to protect the line nodes of band crossing, while the possibilities in lower-dimensional materials are rarely discussed. Here we demonstrate robust NLSM phase in a quasi-one-dimensional nonmagnetic semimetal $TaNiTe_5$. Combining angle-resolved photoemission spectroscopy measurements and first-principles calculations, we reveal how reduced dimension can interact with nonsymmorphic symmetry and result into multiple Dirac-type nodal lines with four-fold degeneracy. Our findings suggest rich physics and application in (quasi-)one-dimensional topological materials and call for further investigation on the interplay between the quantum confinement and nontrivial band topology.

*Introduction*

Symmetry plays an essential role in topological phases of matter because it determines the way in which different wavefunctions are forced to have the same energy eigenvalue, i.e., degeneracy. When the valence and conduction bands cross and form distinct point-contact nodes instead of opening a gap, the system turns to a topological semimetal [1]. Following the concept of particle physics, such elementary excitations could be described as Dirac [2-8] and Weyl fermions [9-20], as well as triply-degenerate [21-24] and double Dirac fermions[25], etc., manifesting novel phenomena such as ultrahigh mobility[8] and unusual magnetic transport behaviors [26,27]. In addition, the possible nodal structure in topological semimetals could also extend to shapes of line, loop and chain in the reciprocal space, leading to the so-called nodal-line semimetals (NLSMs)[28-33]. Typically, NLSMs, especially those robust against spin–orbit coupling (SOC), usually occurs in three-dimensional (3D) materials with more symmetry operations (e.g., mirror and nonsymmorphic symmetries) that protect the Dirac band crossing[34-42]. In comparison, the possibility of robust NLSM phase against SOC in lower-dimensional materials are rarely explored [43-45]. Only very recently, indirect evidence of nonsymmorphic symmetry-protected nodal lines from tunneling spectroscopy was reported based on a two-dimensional (2D) platform, tri-atomic layers bismuth[46]. Direct evidence of robust lower-dimensional NLSMs is still missing.

Here, we report the realization of robust Dirac nodal lines (DNLs), i.e., nodal lines with four-fold degeneracy, in a quasi-one-dimensional (Q1D) nonmagnetic semimetal $TaNiTe_5$. Combining angle-resolved photoemission spectroscopy (ARPES) and density functional theory (DFT) calculations, we reveal that the interplay between nonsymmorphic symmetry and Q1D geometry results into multiple DNLs robust against SOC, as presented by three key features: (1) four-fold degenerate nodes (Dirac cones) at the boundary (Z point) of the bulk Brillouin zone (BZ); (2) these Dirac cones extend through the whole 3D BZ along $T-Z-T$ line and form DNLs; (3) multiple nodal loops in $Z-A-R$ plane. The consistency between ARPES measurements and DFT calculations justifies the above claims and provides direct evidence of the first-discovered Q1D NLSM phase robust against SOC in $TaNiTe_5$. Our results open a door for topological states of matter in Q1D materials and call for further investigation on the interplay between the quantum confinement and nontrivial band topology.

*Lattice structure and characterization*

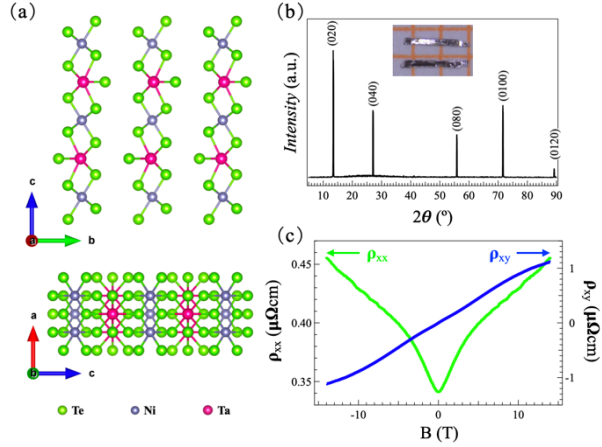

**FIG. 1.** (a) Schematic Q1D lattice structure of TaNiTe$_5$. The single crystals are cleaved along $ac$ plane (bottom panel) for ARPES measurement. (b) Single crystal XRD pattern with all the peaks indexed. The inset shows the optical image of needle-like Q1D single crystals. (c) Field dependent transverse resistivity ($\rho_{xx}$) and Hall resistivity ($\rho_{yx}$) measured at 1.6 K for magnetic field parallel to the $b$ axis and current parallel to the $a$ axis.

Single crystals of TaNiTe$_5$ adapt an orthorhombic layered structure with a space group $Cmcm$ (No. 63)[47-49]. The lattice constants can be inferred from powder X-ray diffraction (XRD) (Fig. S1) and single crystal XRD results (Table S1) with $a = 3.657$ Å, $b = 13.125$ Å and $c = 15.119$ Å. As shown in Fig. 1(a), the one-dimensional NiTe$_2$ chains arrange along the crystallographic $a$ axis and form a quasi-2D layer via linking chains of Ta atoms along the $c$ axis. For ARPES measurement, cleavage occurs parallel to these $ac$ layers ([010]) and normal to the $b$ axis. The crystals grow in the shape of needles along $a$ axis (Fig. 1(b) inset). All the peaks of single crystal XRD pattern in Fig. 1(b) can be indexed as (0 L 0) reflections and no trace of impurity phase can be detected, indicating its high crystalline quality. Field-dependent resistivity measurement was performed with magnetic field parallel to the $b$ axis and current parallel to the $a$ axis. The transverse resistivity $\rho_{xx}$ show linear behavior at low field region ($-4$ T $<$ B $<$ 4 T), indicating possible linear dispersions close to the Fermi level. No saturation can be inferred for field up to 15 T, in contrast to recent report[47]. The Hall resistivity shows generally positive slope with field and clearly non-linear behavior, suggesting multiple Fermi surfaces and hole-type carriers. All these deduced electronic features are confirmed by ARPES data discussed below.

*Multiple Dirac nodal lines in Q1D TaNiTe₅*

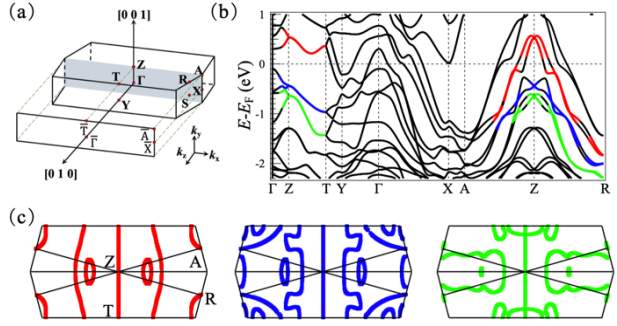

**FIG. 2.** (a) The 3D BZ of bulk TaNiTe$_5$ and the projected BZ for the ARPES measured [010] surface with high-symmetry points specified. (b) DFT calculated electronic band structure of TaNiTe$_5$ along high-symmetry directions with SOC, where the Dirac nodal lines and Dirac cones for different bands near Fermi level are highlighted in red, blue, and green, respectively. (c) The line nodes on $Z-A-R$ plane for different bands, which correspond to red, blue, and green band crosses as shown in (b).

To uncover the topological properties of TaNiTe$_5$, we start from DFT calculation with the lattice constants obtained from our XRD results. Fig. 2(b) shows the calculated band structures of TaNiTe$_5$ with SOC on high-symmetry $k$-paths illustrated in Fig. 2(a). Firstly, it can be clearly seen that the band crossings at the $Z$ point with linear dispersion in $k_x$ ($Z-A$) and $k_y$ ($\Gamma-Z$) directions, forming the typical four-fold degenerate Dirac cones. This is because nonsymmorphic symmetry $g'_y = \{M_y|(0,\frac{1}{2},0)\}$ and $S_{2y} = \{C_{2y}|(0,\frac{1}{2},0)\}$ fulfills the anti-commutation relationship with inversion at Z, leading to an extra two-fold degeneracy between two pairs of Kramers doublets. Secondly, along the $k_z$ direction, the Dirac cones extend from point $Z$ to point $T$, resulting into robust DNLs against strong SOC. These Dirac cones and DNLs features are highlighted in red, blue, and green for the bands near the Fermi level along $\Gamma-Z-T$ path shown in Fig. 2(b). Thirdly, the band width along $k_x$ ($A-Z$) direction is around 1.5 eV, almost 5 times larger than that along $k_y$ ($\Gamma-Z$) and $k_z$ ($T-Z$), reflecting its Q1D character of TaNiTe$_5$. Interestingly, these strongly dispersed bands along $A-Z-R$ path appear in pairs, forming Type-I Dirac cones on point $Z$, Type-II Dirac cones on point $R$, and cross each other several times forming Type-II Dirac cones in between, which are also highlighted in colors in Fig. 2(b). Further DFT calculation confirms that all Dirac cones connect each other forming nodal lines and nodal loops on $k_y = \frac{\pi}{b}$ ($Z-A-R$) plane, as shown in Fig. 2(c). We prove that these DNLs are protected by nonsymmorphic symmetry in the presence of strong SOC in TaNiTe$_5$, providing evidence of robust NLSM phase in a Q1D system (see Supplementary materials for details).

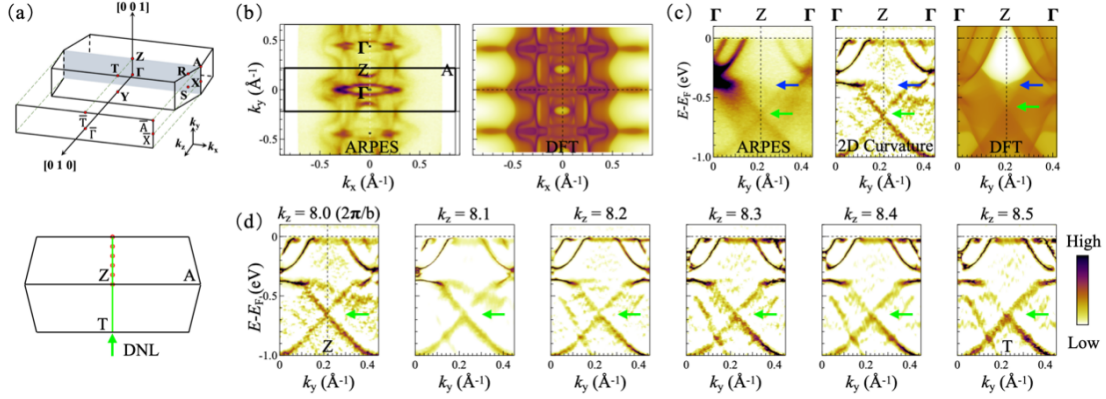

**FIG. 3.** (a) Schematic 3D BZ and the location of the DNLs along T − Z − T line in the top Z − A − T plane. Red open circles between Z and T indicate the positions where ARPES spectra in (d) are taken. (b) Comparison between Fermi surfaces from ARPES (left) and DFT (right) for single crystal $ac$ plane (Γ − Z − A plane) of TaNiTe$_5$. (c) Comparison between dispersions along Γ − Z − Γ from ARPES raw spectra (left), 2D curvature (middle) and DFT (right). (d) Photon energy dependent dispersion along $\bar{\Gamma} - \bar{T} - \bar{\Gamma}$ for $k_z$ covering half of the 3D BZ from Z to T.

Our ARPES measurements provide direct evidence of these multiple DNL features. We first analyze the Dirac cone at bulk Z point and DNLs along T − Z − T line as indicated in Fig. 3(a). Fermi surface and bands measured by ARPES and projected from DFT calculation are shown in comparison in Fig. 3(b, c). For ARPES measurement, the photon energy was selected as 50 eV, so it approximately covers the Γ − Z − A plane, see Fig. S2 for systematic photon energy-dependent data. Satisfactory agreement between experiment and theory is reached considering the detailed Fermiology and spectrum. In particular, the four-fold Dirac cone crossing at the bulk Z point (as discussed in Fig. 2) can be clearly resolved by ARPES, as marked in Fig. 3(c) by green arrows. This measured Dirac cone corresponds to the calculated one shown in green lines in Fig. 2(b). The calculated Dirac cone shown in red lines is above the Fermi level and the one in blue lines is too weak to be resolved (blue arrows in Fig. 3(c)).

Judging from the highlighted dispersion in the curvature plot, this Dirac crossing is gapless at Z point. Furthermore, in Fig. 3(d) we present the spectra of this Dirac crossing at different $k_z$ planes measured by varying the incident photon energy. It is clear that this Dirac crossing remains gapless across the 3D BZ, composing a DNL from Z to T point as indicated by the green curves in Fig. 2(c) and Fig. 3(a). It is noted that this DNL shows clear dispersion along $Z-T$ line in DFT (Fig. 2(b)) but is quite flat from ARPES, likely attributed to the $k_z$ broadening of low-energy ARPES and the functional chosen in DFT to estimate the interlayer coupling. Nevertheless, photon energy dependent ARPES measurement corroborates DFT calculation and establishes a gapless Dirac cone at bulk Z point and a gapless DNL along $T-Z-T$ line in the 3D BZ.

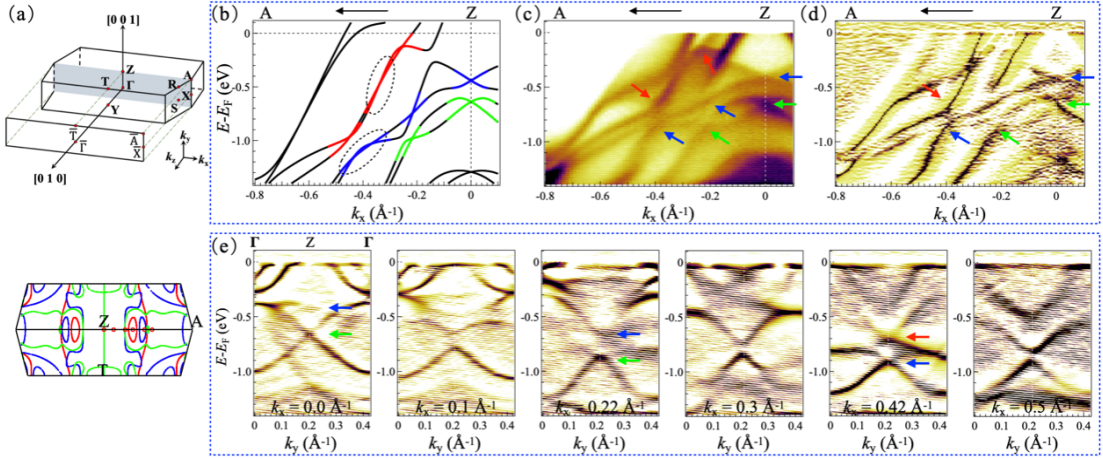

**FIG. 4.** (a) Schematic 3D BZ and the location of the nodal loops in $T-Z-A$ plane. Red solid circles between Z and A indicate the positions where ARPES spectra in (e) are taken. (b, c, d) Comparison between spectra along $Z \rightarrow A$ direction from DFT bands (b), ARPES raw spectra (c) and 2D curvature ARPES spectra (d). (e) Systematic evolution of spectra (2D curvature) along $k_y$ at different $k_x$ values. Colored arrows in (c, d and e) indicate the Dirac crosses from the corresponding band pairs shown in the DFT calculation (b).

Considering the nodal loops locating in $T-Z-A$ plane, we can also find strong evidence of their corresponding Dirac crosses from the satisfactory match between ARPES and DFT. In Fig. 4(a) bottom panel, the nodal loops from three pairs of bands close to the Fermi level are shown in $T-Z-A$ plane with red, blue, green colors as in Fig. 2(c). ARPES spectra and DFT dispersion along $k_x$ (Z → A) direction are shown in comparison in Fig. 4(b, c, d). In the present DFT energy-momentum window (Fig. 4(b)), two Dirac crosses can be identified for the green pair of bands, four Dirac crosses for the blue pair and three for the red, with all the band pairs containing both type I and type II Dirac cones. The dashed ellipses in Fig. 4(b) point out regions where the Dirac gaps are too small to be resolved in ARPES spectra. Except that, all the features predicted by DFT can be identified in the ARPES spectra, as emphasized in Fig. 4(c) and 4(d)by arrows colored correspondingly. The existence of multiple type I and type II Dirac crosses along $k_x$ (Z → A) direction provides strong evidence for the nodal loops shown in Fig. 4(a).

Further evidence of these Dirac crosses can also be derived from the ARPES spectra along $k_y$ (parallel to Γ − Z − Γ) direction at different values of $k_x$. As shown in Fig. 4(e), ARPES spectra along $k_y$ direction at different $k_x$ positions present evolution of the Dirac crosses corresponding to the nodal loop shape. When the $k_x$ positions of the spectra, as specified by red circles in Fig. 4(a) bottom panel, coincide with the nodal loops, gapless Dirac crosses can be identified directly in the 2D curvature spectra. These Dirac crosses are also emphasized by arrows whose colors correspond to the band pairs. When the $k_x$ positions are off the nodal loops, gaps open for the Dirac cone. The correspondence between ARPES and DFT down to very detailed level establishes solid evidence of multiple Dirac nodal loops in T − Z − A plane.

*Summary*

Combining ARPES measurements, DFT calculations and symmetry analysis, we have demonstrated how reduced dimension can interact with nonsymmorphic symmetry and reshape the band topology in a Q1D semimetal. The resulting multiple robust Dirac crosses and DNLs against SOC suggest fertile physics in lower-dimensional topological systems[50-52]. Compared to their 3D and 2D counterparts, Q1D topological materials could be better suited for device exploitation due to their reduced dimensionality, especially in devices utilizing coherent spin transport[53,54]. Our work would stimulate further effort in exploring these rich physics and applications in Q1D topological materials.

**ACKNOWLEDGEMENTS**

This work is supported by National Key R&D Program of China (Grants No. 2020YFA0308900), the National Natural Science Foundation of China (NSFC) (Grants No. 12074163, No. 12074161), NSFC Guangdong (No. 2016A030313650), the Shenzhen High-level Special Fund (Grants No. G02206304 and No. G02206404), the Guangdong Innovative and Entrepreneurial Research Team Program (Grants No. 2019ZT08C044), Shenzhen Science and Technology Program (Grant No. KQTD20190929173815000), the University Innovative Team in Guangdong Province (No. 2020KCXTD001), Guangdong Provincial Key Laboratory for Computational Science and Material Design under Grant No. 2019B030301001, the Science, Technology and Innovation Commission of Shenzhen Municipality (No. KQTD20190929173815000) and Center for Computational Science and Engineering of Southern University of Science and Technology.